\documentclass{PoS}

\title{Simulation of jet quenching and high-p$_T$ particle production 
at RHIC and LHC}

\ShortTitle{Simulation of jet quenching and high-p$_T$ particle production 
at RHIC and LHC}

\author{\speaker{I.P. Lokhtin}, S.V. Petrushanko, A.M. Snigirev, 
C.Yu. Teplov\\
        Skobeltsyn Institute of Nuclear Physics, Lomonosov Moscow State
 University\\
        E-mail: \email{Igor.Lokhtin@cern.ch}, 
 \email{sergant@lav01.sinp.msu.ru}, 
 \email{snigirev@lav01.sinp.msu.ru}, 
 \email{teplov@lav01.sinp.msu.ru}}

\abstract{The model to simulate rescattering and partonic energy loss in 
ultrarelativistic heavy ion collisions is presented. The full heavy ion event 
is obtained as a superposition of a soft hydro-type state and hard multi-jets. 
This model is capable of reproducing main features of the jet quenching pattern 
at RHIC, and is applied to probe jet quenching in various novel channels at 
LHC.} 

\FullConference{High-pT physics at LHC\\
   March 23-27 2007\\
   University of Jyvaskyla, Jyvaskyla, Finland}

\begin{document}

\section{Introduction}

One of the important tools for studying the properties of quark-gluon plasma 
(QGP) in ultrarelativistic heavy ion collisions is the analysis of a QCD jet 
production. The medium-induced energy loss of energetic partons, ``jet 
quenching'', should be very different in the cold nuclear matter and QGP, 
resulting in many observable phenomena~\cite{Baier:2000m}. Recent RHIC data on 
high-p$_T$ particle production at $\sqrt{s}=200~A$ GeV are in agreement with 
the jet quenching hypothesis (see, e.g.,~\cite{Wang:2003aw} and references 
therein). At LHC, a new regime of heavy ion physics will be reached at 
$\sqrt{s_{\rm NN}}=5.5 A$ TeV where hard and semi-hard particle production can 
stand out against the underlying soft events. The initial gluon densities in 
PbPb reactions at LHC are expected to be much higher than those at RHIC, 
implying a stronger partonic energy loss, observable in new channels.

In the most of available Monte-Carlo heavy ion event generators the 
medium-induced partonic rescattering and energy loss are either ignored or 
implemented insufficiently. Thus, in order to analyze RHIC data on high-p$_T$ 
hadron production and test the sensitivity of LHC observables to the QGP 
formation, the development of adequate and fast Monte-Carlo tool to simulate the
jet quenching is necessary.  

\section{Physics model and simulation procedure} 

The detailed description of physics model can be found in our recent 
paper~\cite{Lokhtin:2005px}. The approach bases on an accumulating energy 
loss, the gluon radiation being associated with each parton scattering in the 
expanding medium and includes the interference effect using the modified 
radiation spectrum $dE/dl$ as a function of decreasing temperature $T$. The 
basic kinetic integral equation for the energy loss $\Delta E$ as a function of 
initial energy $E$ and path length $L$ has the form 
\begin{equation}
\label{elos_kin}
\Delta E (L,E) = \int\limits_0^L dl \frac{dP(l)}{dl}\lambda (l)
\frac{dE(l,E)}{dl},~~~~
\frac{dP(l)}{dl} = \frac{1}{\lambda (l)} \exp{(-l/\lambda (l))},
\end{equation}
where $l$ is the current transverse coordinate of a parton, $dP/dl$ is the 
scattering probability density, $dE/dl$ is the energy loss per unit length, 
$\lambda$ is in-medium mean free path. The collisional loss in high-momentum 
transfer limit and radiative loss in BDMS approximation~\cite{Baier:1999ds} 
(with ``dead-cone'' generalization of the radiation spectrum for heavy 
quarks~\cite{Dokshitzer:2001zm}) are using. We consider realistic nuclear 
geometry and treat the medium in nuclear overlapping zone as a 
boost-invariant longitudinally expanding quark-gluon fluid. The model 
parameters are the initial conditions for the QGP formation for central AuAu 
(PbPb) collisions at RHIC (LHC): the proper formation time $\tau_0$ and 
the temperature $T_0$. For non-central collisions we suggest the 
proportionality of the initial energy density $\varepsilon _0$ to the ratio of 
nuclear overlap function and transverse area of nuclear 
overlapping. The simple Gaussian parameterization of gluon 
angular distribution over the emission angle $\theta$ with the typical angle of 
the coherent radiation $\theta _0 \sim 5^0$~\cite{Lokhtin:1998ya} is used. 

The model was constructed as the Monte-Carlo event generator PYQUEN 
(PYthia QUENched) and is available via Internet~\cite{pyquen}. The routine is 
implemented as a modification of the standard PYTHIA$\_$6.4 jet 
event~\cite{Sjostrand:2000wi}. The event-by-event simulation procedure 
includes the generation of the initial parton spectra with PYTHIA and 
production vertexes at given impact parameter, rescattering-by-rescattering 
simulation of the parton path length in a dense zone, radiative and collisional 
energy loss per rescattering, final hadronization with the Lund string model 
for hard partons and in-medium emitted gluons.

The full heavy ion event is simulated as a superposition of soft hydro-type 
state and hard multi-jets. The simple approximation~\cite{Lokhtin:2005px} of 
hadronic liquid at ``freeze-out'' stage has been used to treat soft part of the 
event. Then the hard part of the event includes PYQUEN multi-jets generated 
according to the binomial distribution. The mean number of jets produced in AA 
events at a given impact parameter is a product of the number of binary NN 
sub-collisions and the integral cross section of hard process in $pp$ 
collisions with the minimal transverse momentum transfer $p_T^{\rm min}$. The 
extended in such a way model has been also constructed as the fast Monte-Carlo 
event generator~\cite{hydjet}. Note that ideologically similar approximation 
has been developed in~\cite{Hirano:2004rs}. 

\section{Jet quenching at RHIC}

\begin{figure}
\begin{minipage}{17pc}
\includegraphics[width=17pc]{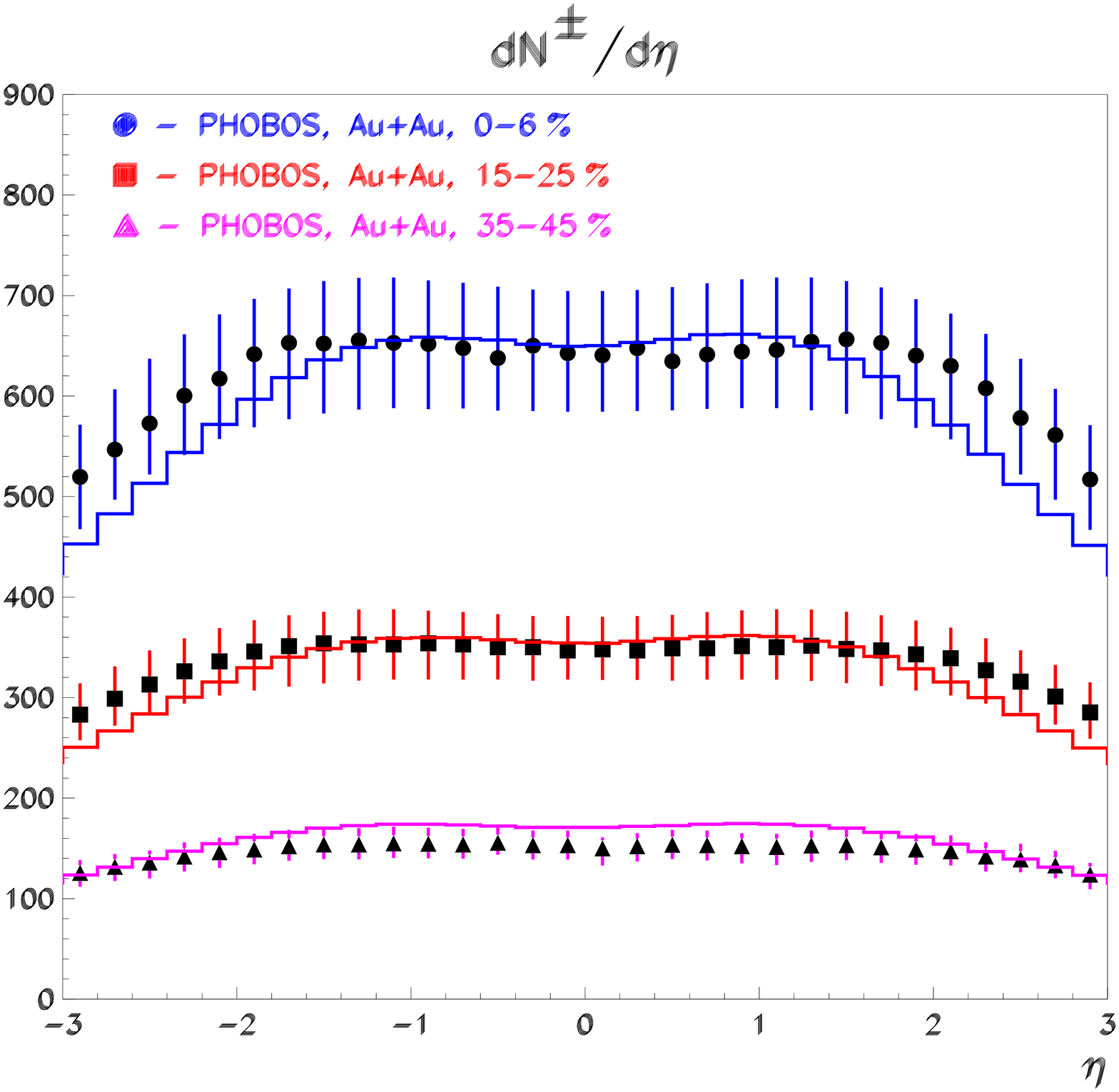}
\caption{The pseudorapidity distribution of charged hadrons in AuAu 
collisions for three centrality sets. The points are PHOBOS data, 
histograms are the model calculations.}
\label{fig1}
\end{minipage}
\hspace{\fill}%
\begin{minipage}{17pc}
\includegraphics[width=17pc]{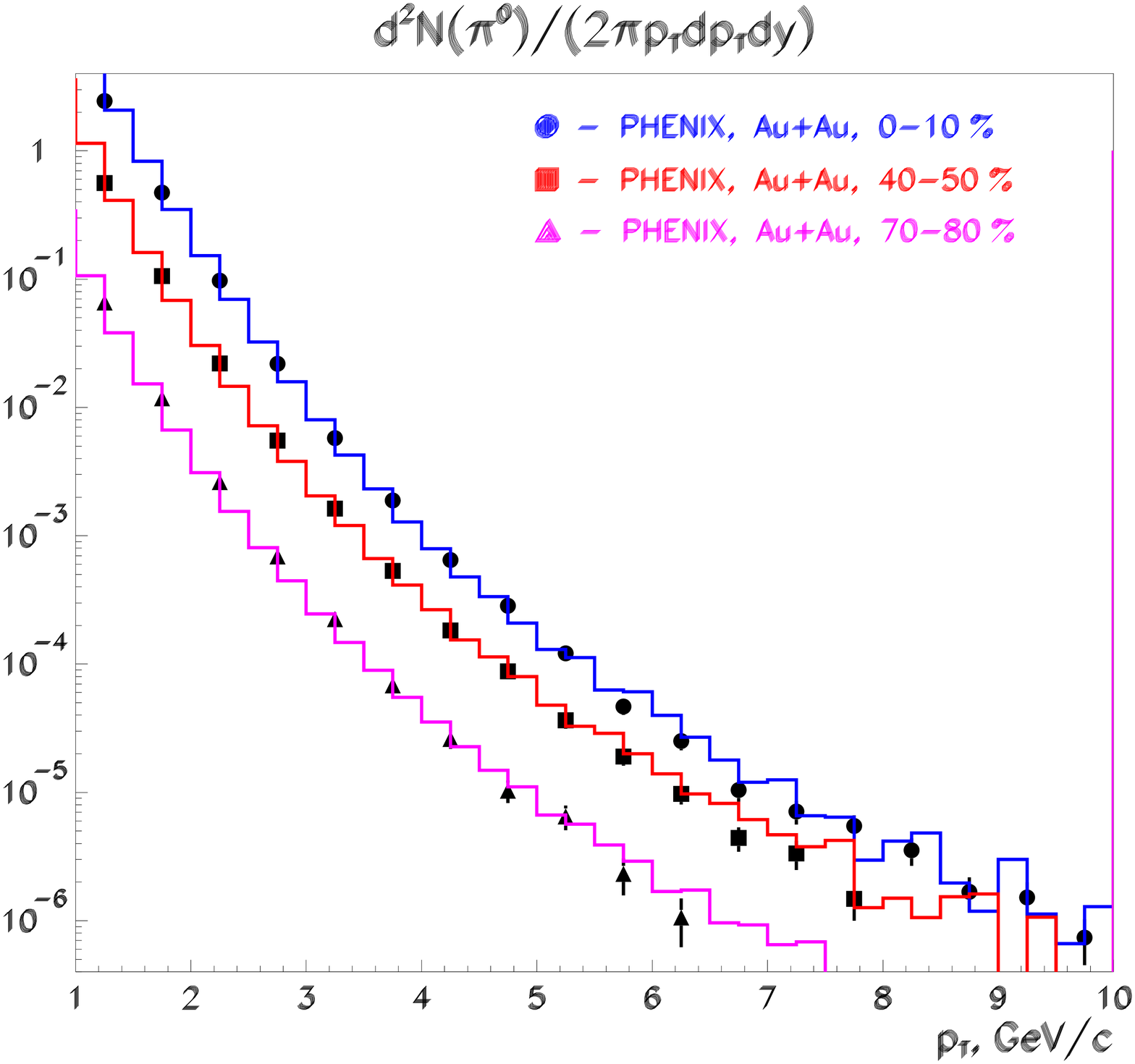}
\caption{The transverse momentum distribution of neutral pions in AuAu 
collisions for three centrality sets. The points are PHENIX data, 
histograms are the model calculations.}
\label{fig2}
\end{minipage}
\end{figure}

\begin{figure}
\begin{minipage}{17pc}
\includegraphics[width=17pc]{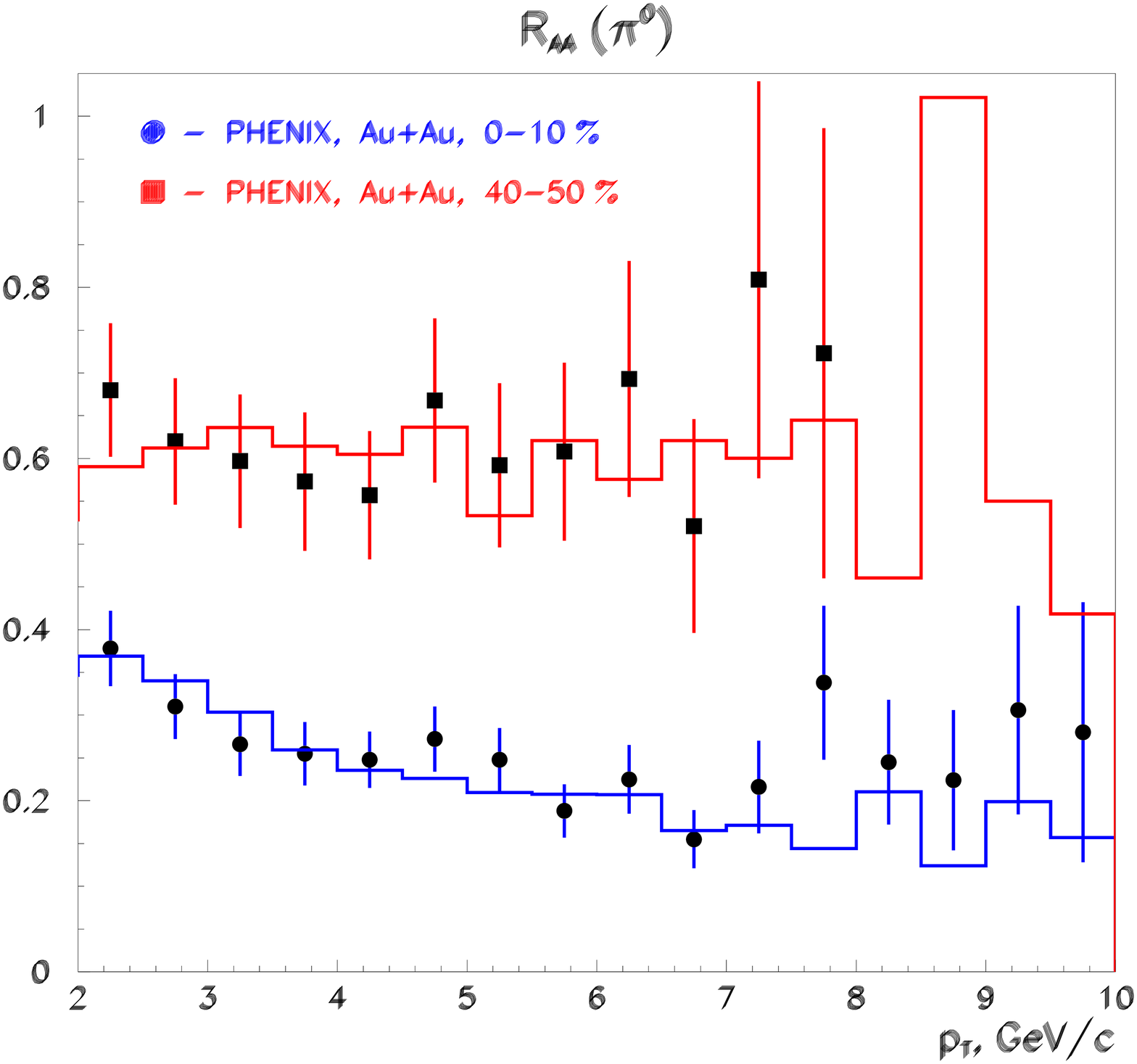}
\caption{The nuclear modification factor $R_{AA}$ for 
neutral pions in AuAu collisions for two centrality sets. The points are  
PHENIX data, histograms are the model calculations.}
\label{fig3}
\end{minipage}
\hspace{\fill}%
\begin{minipage}{17pc}
\includegraphics[width=17pc]{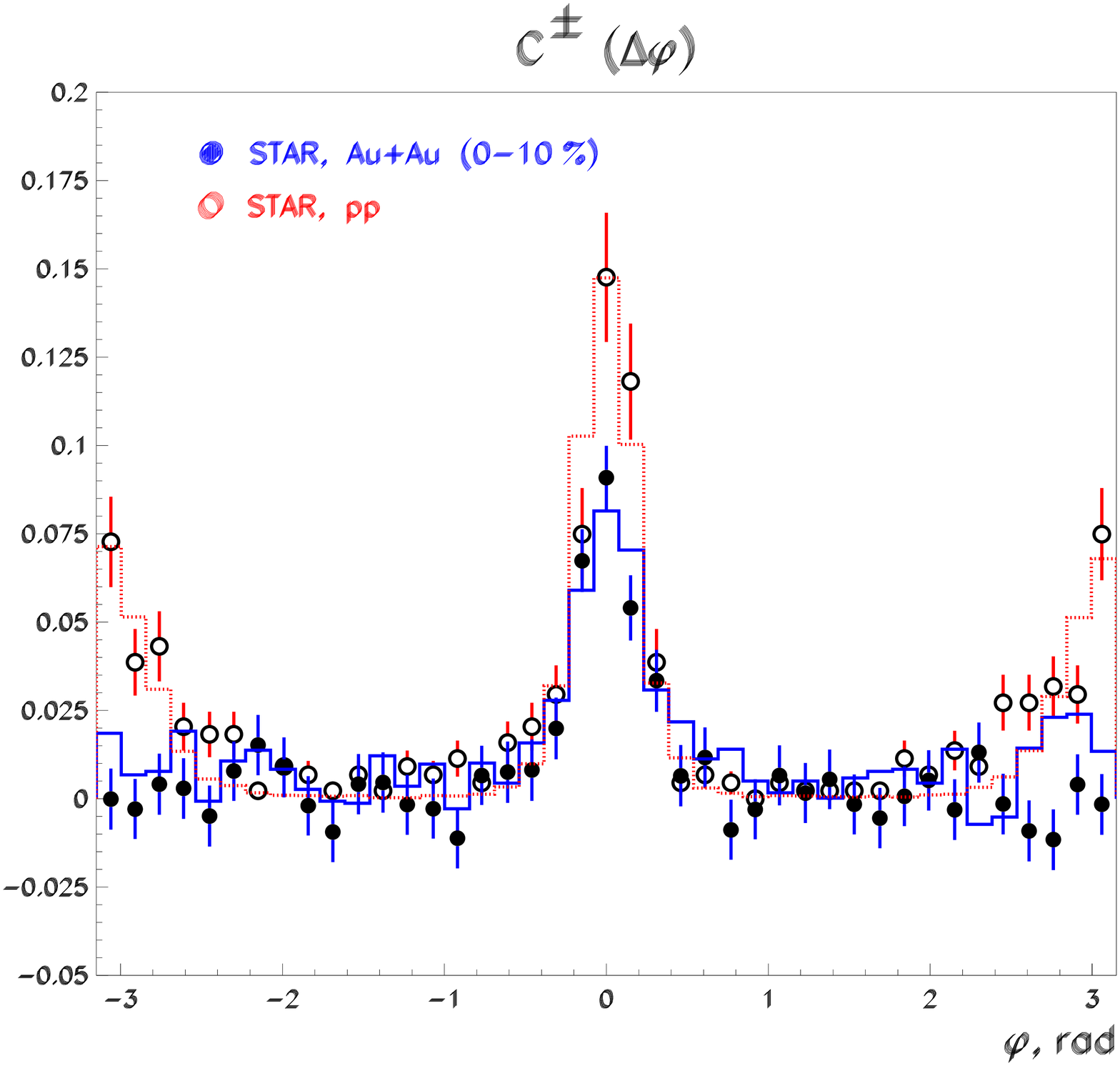}
\caption{The azimuthal two-particle correlation function for pp and for 
central AuAu collisions. The points are STAR data, dashed and 
solid histograms are the model calculations for pp and AuAu events 
respectively.}
\label{fig4}
\end{minipage}
\end{figure}

In order to demonstrate the efficiency of the model, the jet quenching pattern 
in AuAu collisions at RHIC was considered. The PHOBOS data on $\eta$-spectra 
of charged hadrons~\cite{Back:2002wb} have been analyzed at first to fix the 
particle density in the mid-rapidity and the maximum longitudinal flow 
rapidity, $Y_L^{\rm{max}}=3.5$ (figure \ref{fig1}). The rest of the model 
parameters were obtained by fitting PHENIX data on $p_T$-spectra of neutral 
pions~\cite{Adler:2003qi} (figure \ref{fig2}): the kinetic freeze-out 
temperature $T_f=100$ MeV, maximum transverse flow rapidity 
$Y_T^{\rm{max}}=1.25$ and minimum transverse momentum of hard parton-parton 
scattering $p_T^{\rm min}=2.8$ GeV/$c$. The nuclear 
modification of the hardest domain of $p_T$-spectrum was used to extract 
initial QGP conditions: $T_0=500$ MeV and $\tau_ 0=0.4$ fm/$c$. Figure 
\ref{fig3} shows that our model well reproduces $p_T$-- and centrality 
dependences of nuclear modification factor $R_{AA}$, which is defined  as:
$$R_{\rm AA}(p_T,\eta;b)=\frac{\sigma_{\rm pp}^{\rm inel}}
{\langle N_{\rm coll}\rangle}\frac{d^2N_{AA}/d p_Td\eta}
{d^2\sigma_{\rm pp}/d p_Td\eta}~,$$ where 
$\langle N_{\rm coll}\rangle = T_{\rm AA}(b) \times \sigma_{\rm pp}^{\rm inel}$ 
is the average number of binary nucleon-nucleon collisions in a given impact
parameter b (with nucleus overlap function $T_{\rm AA}(b)$). If there are no 
nuclear effects, the value of $R_{\rm AA}$ at high $p_T$ should be unity.

Another important tool to verify jet quenching is two-particle azimuthal 
correlation function $C(\Delta \varphi)$ -- the distribution over an azimuthal 
angle of high-$p_T$ hadrons in the event with $2$ GeV/$c<p_T<p_T^{\rm trig}$  
relative to that for the hardest ``trigger'' particle with $p_T^{\rm trig}>4$ 
GeV/$c$. Figure \ref{fig4} presents $C(\Delta \varphi)$ in pp and in central 
AuAu collisions (data from STAR~\cite{Adler:2002xw}). Clear peaks in pp 
collisions at $\Delta \varphi = 0$ and $\Delta \varphi =\pi$ indicate a typical 
dijet event topology. However, for central AuAu collisions the peak near $\pi$ 
disappears. It can be interpreted as the observation of monojet events due to 
the absorption of one of the jets in a dense medium. Figure 4 demonstrates that 
measured suppression of azimuthal back-to-back correlations is well reproduced 
by our model.  

We leave beyond the scope of this paper the analysis of such important RHIC 
observables as the azimuthal anisotropy and particle ratios at low $p_T$. In 
order to study them, a more careful treatment of soft particle production than 
our simple approach is needed (the detailed description of space-time structure 
of freeze-out region, resonance decays, etc.). 

\section{Jet quenching at LHC}

The developed model was applied to analyze various novel features of jet
quenching in heavy ion collisions at the LHC. Let us give a few examples 
of such jet observables. All calculations have been done for PbPb collisions 
at $\sqrt{s_{\rm NN}} = 5.5$ TeV with PYQUEN energy loss model. The jet was 
defined on the generator level by a simple way, just collecting the energy 
around the direction of a leading particle inside a cone 
$R\,=\,\sqrt{\Delta \eta^2+\Delta \varphi ^2}=0.5$. The pseudorapidity cuts 
corresponding to the geometrical acceptance of CMS experiment were applied: 
$|\eta|<3$ for jets and neutral hadrons, $|\eta|<2.5$ for charged hadrons and 
muons. 

\subsection{Nuclear modification factors for jets} 

The nuclear modification factor can be determined for jets by the same way as 
for inclusive hadron production. Since at the LHC no pp data will be available 
at $\sqrt{s_{\rm NN}}$ = 5.5 TeV at the time of the first PbPb data taking, 
particle spectra in pp collisions will be interpolated to this energy using 
perturbative QCD predictions constrained by the existing Tevatron data at 1.8 
TeV and by the LHC results at 14 TeV. Another possibility to quantifies 
medium-modified particle spectra is to use the central to peripheral heavy ion 
collision ratio, $R_{\rm CP}(p_T,\eta)$, which does not require a pp reference, 
but has rather limited statistical reach of the peripheral data set.

\begin{figure}
\begin{minipage}{17pc}
\includegraphics[width=17pc]{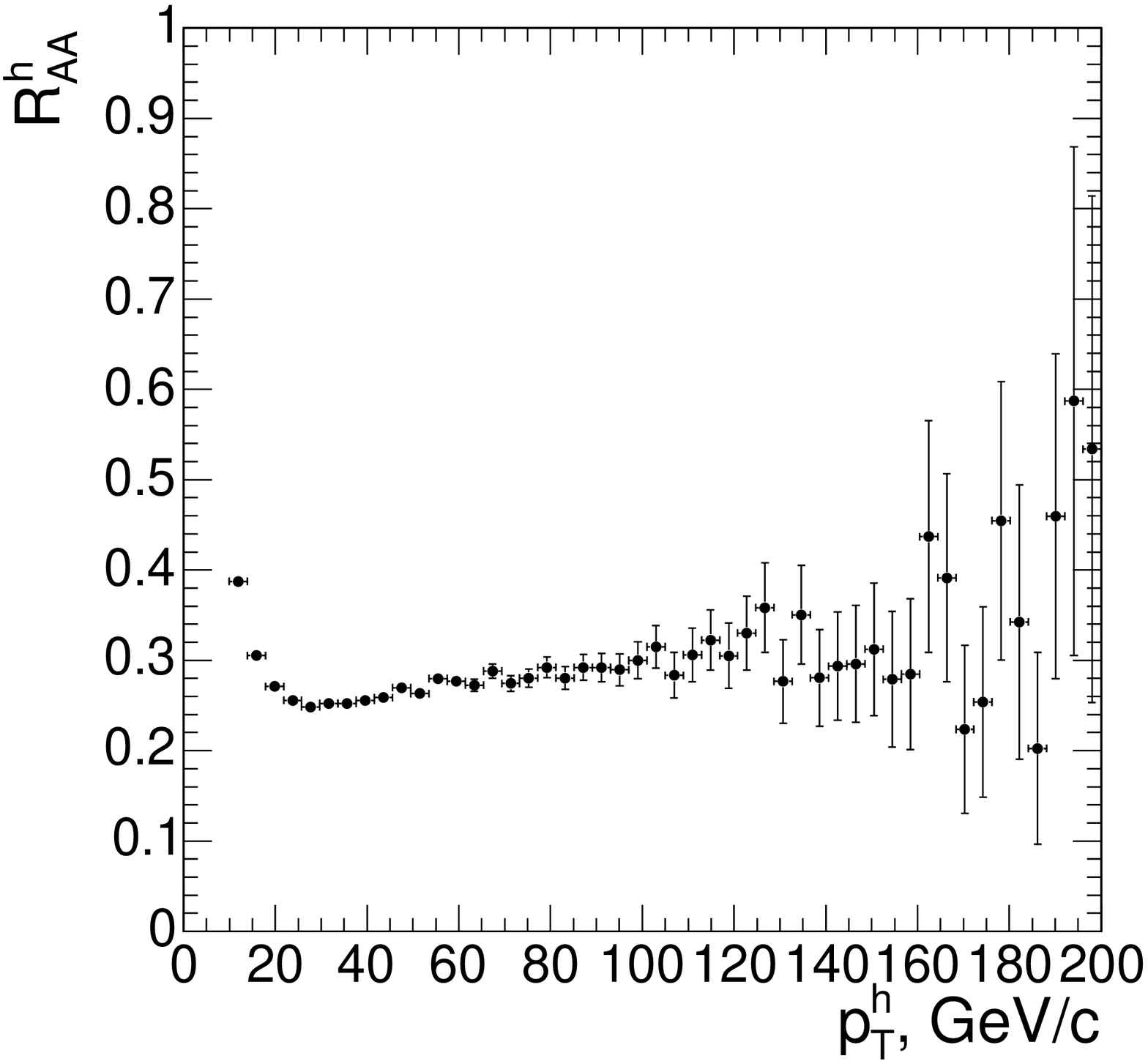}
\caption{The nuclear modification factor, $R^{\rm h}_{\rm AA}(p_T)$, for 
inclusive charged hadrons in central PbPb collisions triggered on jets with 
$E_T^{\rm jet}>100$ GeV. The number of histogram entries and statistical errors 
correspond to the estimated event rate for one month of LHC run. 
\label{fig:raa_had}}
\end{minipage}
\hspace{\fill}%
\begin{minipage}{17pc}
\includegraphics[width=17pc]{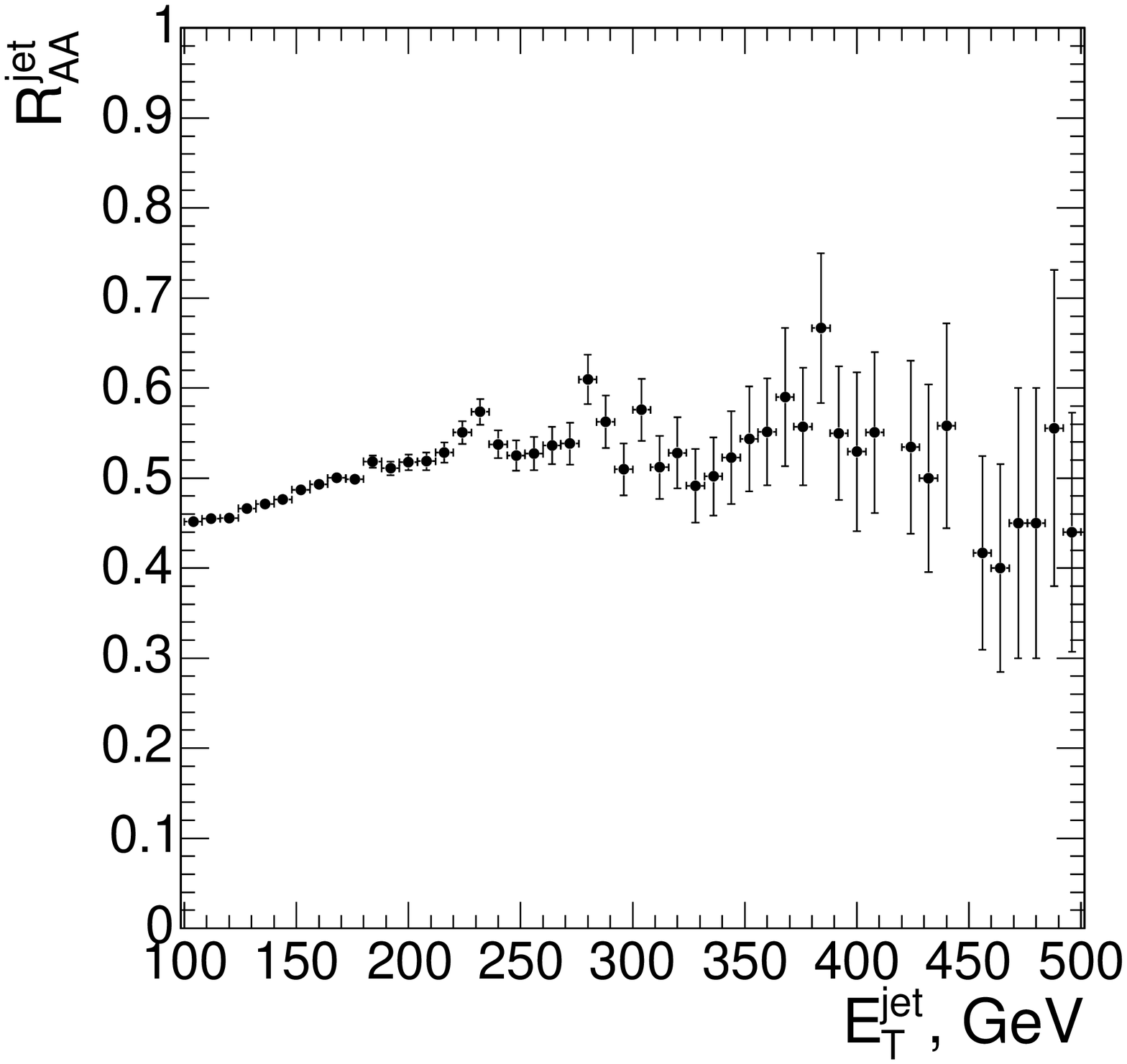}
\caption{The nuclear modification factor, $R^{\rm jet}_{\rm AA}(p_T)$, for 
jets of cone size $R=0.5$ in central PbPb collisions. The number of histogram 
entries and statistical errors correspond to the estimated event rate for one 
month of LHC run. 
\label{fig:raa_jet} }
\end{minipage} 
\end{figure}

Figure~\ref{fig:raa_had} shows the $p_T$-dependence of nuclear modification 
factor, $R^{\rm h}_{\rm AA}(p_T)$, for inclusive charged hadrons in central 
PbPb events triggered on jets with $E_T^{\rm jet}>100$ GeV. The number of 
entries and the statistical errors correspond to the 
estimated event rate for one month of LHC run and a nominal integrated 
luminosity of 0.5 nb$^{-1}$~\cite{Virdee:1019832}. The estimated suppression 
factor slightly increases with $p_T$($>20$ GeV), from $\sim 0.25$ at 
$p_T \sim 20$ GeV to $\sim 0.35$ at $p_T \sim 200$ GeV. This behaviour 
manifests the specific implementation of partonic energy loss in the model, 
rather weak energy dependence of loss and the shape of initial parton spectra. 
Without event triggering on high-$E_T$ jet(s), the suppression is 
stronger ($\sim 0.15$ at 20 GeV and slightly increasing with $p_T$ up to 
$\sim 0.3$ at 200 GeV). 

A novel observable at the LHC will be the nuclear modification factor for hard 
jets, which can be reconstructed in high multiplicity environment with a good 
efficiency and low background starting from the energy 
$E_T^{\rm jet} \sim 50-100$ GeV~\cite{Virdee:1019832}. 
Figure~\ref{fig:raa_jet} shows the $p_T$-dependence of jet nuclear 
modification factor, $R^{\rm jet}_{\rm AA}(p_T)$. The other conditions are the 
same as it was described above. The estimated suppression factor (due to 
partial gluon bremsstrahlung out of jet cone and collisional loss) is about 
$2$ and almost independent on jet energy. The measured jet nuclear 
modification factor will be very sensitive to the fraction of partonic 
energy loss carried out of the jet cone. 

\subsection{Medium-modified jet fragmentation function}

The ``jet fragmentation function'' (JFF), $D(z)$, is defined as the probability 
for a given product of the jet fragmentation to carry a fraction $z$ of the 
jet transverse energy. In nuclear (AA) interactions, the JFF for 
leading hadrons  (i.e.\ the hadron carrying the largest fraction of the 
jet momentum) can be written as~\cite{Lokhtin:2003yq,Vardanian:2005th}:
\begin{eqnarray}
\label{dz} 
\hskip -2cm 
D(z) = \int \limits_{z\cdot p_{\rm T~{\rm min}}^{\rm jet}} d(p^h_{\rm
T})^2 d y d z'
\frac{d N_{\rm AA}^{\rm h(k)}}{d(p^h_{\rm T})^2d y d z'} \delta
\left( z-\frac{p^h_{\rm T}}{p^{\rm
jet}_{\rm T}} \right) \Bigg/ \int \limits_{p_{\rm T~{\rm
min}}^{\rm jet}} d(p_{\rm T}^{\rm jet})^2 d y
\frac{dN_{\rm AA}^{\rm jet(k)}}{d(p_{\rm T}^{\rm jet})^2d y}~,
\end{eqnarray}
where  $p^h_T \equiv z p_{\rm T}^{\rm jet}=z'p_{\rm T}$ is the transverse 
momentum of a leading hadron, $z'$ is the hadron momentum fraction relative to 
the $p_{\rm T}$ of the parent parton, $p_{T~{\rm min}}^{\rm jet}$ is the 
minimum momentum threshold of observable jets, 
$(d N_{\rm AA}^{\rm jet(k)})/(d(p_{\rm T}^{\rm jet})^2 d y)$
and $(d N_{\rm AA}^{\rm h(k)})/(d(p_{\rm T}^{\rm h})^2d y d z')$ are the 
yields of $k$-type jets and hard hadrons, respectively.

\begin{figure}
\begin{minipage}{17pc}
\includegraphics[width=17pc]{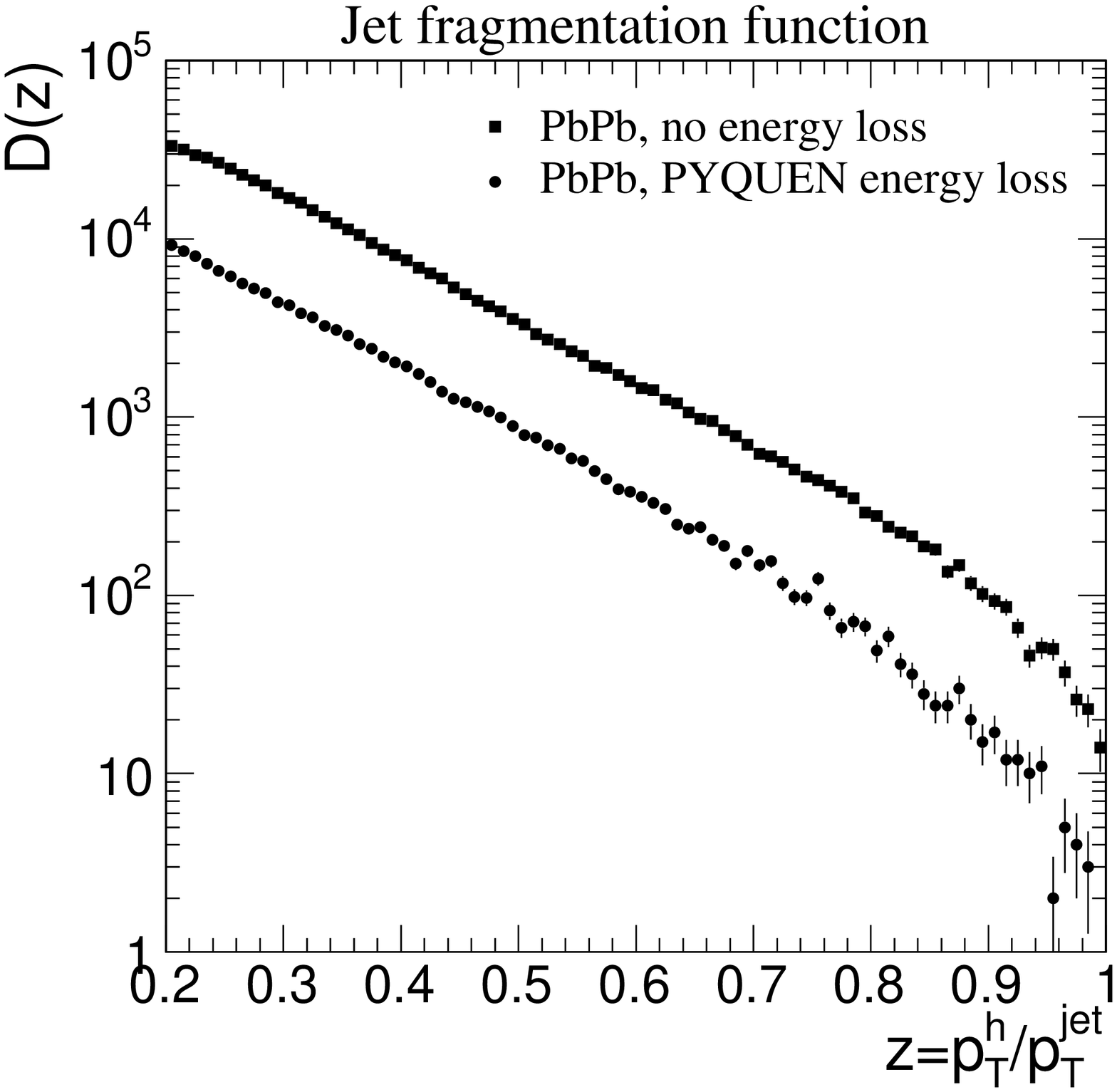}
\caption{Jet fragmentation function for leading hadrons  
($E^{\rm jet}_{\rm T}>100$ GeV)
    in central PbPb  collisions without (squares) and
    with (circles) partonic energy loss. The number of histogram entries and
    statistical errors correspond to the estimated jet rate for one month of 
    LHC run. 
 \label{fig:jff}}
\end{minipage}
\hspace{\fill}%
\begin{minipage}{17pc}

\vskip -0.8 cm

\includegraphics[width=17pc]{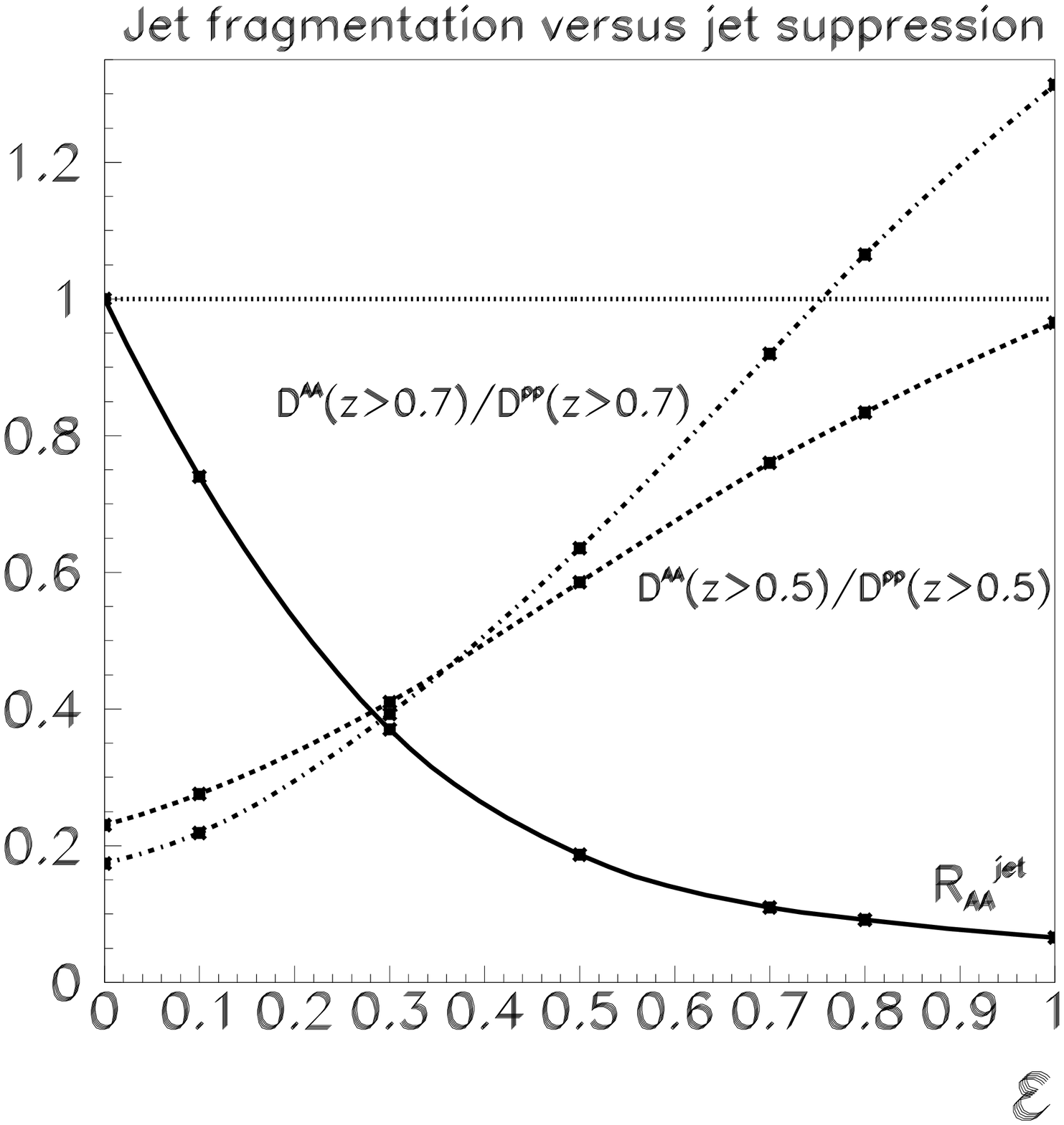}
\caption{Jet nuclear modification factor (solid curve) and 
ratio of JFF with energy loss to JFF without loss ($z>0.5$  for dashed curve 
and $z>0.7$ for dash-dotted curve) as a function of the fraction $\varepsilon$ 
of partonic energy loss carried out of the jet cone.  
\label{fig:jff_vs_quen} }
\end{minipage} 
\end{figure}

Figure~\ref{fig:jff} shows JFF's in central PbPb collisions with and without 
partonic energy loss for $E_{\rm T}^{\rm jet}>100$ GeV. The 
number of entries and the statistical errors again correspond to the estimated 
event rate for one month of LHC run. Significant softening of the JFF (by a 
factor of $\sim 4$ and slightly increasing with $z$) is predicted. 

The medium-modified JFF is sensitive to a fraction $\varepsilon$ of partonic 
energy loss carried out of the jet cone, which is related also to the 
suppression of the absolute jet rates. Figure~\ref{fig:jff_vs_quen} shows the 
$\varepsilon$-dependences of jet nuclear modification factor $R_{\rm AA}^{\rm 
jet}$ and ratio of JFF with energy loss to JFF without loss, 
$D^{\rm AA} (z>z_0) / D^{\rm pp} (z>z_0)$, for $z_0=0.5$ and $0.7$ in central 
PbPb collisions~\cite{Lokhtin:2003yq,Vardanian:2005th}. If $\varepsilon$ close 
to 0, then $R_{\rm AA}^{\rm jet} \sim 1$ (there is no jet rate suppression), 
and JFF softening is maximal. Increasing $\varepsilon$ results in stronger jet 
rate suppression, but effect on JFF softening becomes smaller, especially for 
highest $z$ (the ratio $D^{\rm AA}/D^{\rm pp}$ can be even greater than $1$ at 
large enough $\varepsilon$ and $z$ values). The physical reason for the effect 
to be opposite in the jet suppression factor and the fragmentation function is 
it follows. Increasing $\varepsilon$ results in decreasing final jet transverse 
momentum (which is the denominator in definition of $z$ in JFF~(\ref{dz})) 
without an influence on the numerator of $z$ and, as a consequence, in reducing 
effect on JFF softening, while the integral jet suppression factor becomes 
larger. The crossing point between two effects is $\varepsilon \sim 0.3$.  

Thus a novel concurrent study of the possible softening of the JFF and 
suppression of the absolute jet rates can be carried out in order to 
differentiate between various energy loss mechanisms. Strong JFF softening 
without substantial jet rate suppression would be an indication of small-angle 
gluon radiation dominating the medium-induced partonic energy loss. Increasing  the contribution from 
wide-angle gluon radiation and collisional energy loss leads to jet rate 
suppression with less pronounced softening of the JFF. If, instead, the 
contribution of the ``out-of-cone'' jet energy loss is large enough, the jet 
rate suppression may be even more significant than the JFF softening.

\subsection{Jet azimuthal anisotropy} 

The azimuthal anisotropy of particle spectrum is one of the most important 
tools to study properties of dense QCD-matter created in heavy ion collisions. 
It is usually characterized by the second coefficient of the Fourier 
expansion of particle azimuthal distribution, so called elliptic flow 
coefficient, $v_2$. The momentum dependence of $v_2$ for high-$p_T$ hadrons, 
observed in semi-central AuAu collisions at RHIC, strongly supports the 
presence of rescattering and energy loss of hard partons in the azimuthally 
asymmetric volume of the nuclear reaction. A novel observable at the LHC 
will be the azimuthal anisotropy for hard jets (due to the part of partonic 
energy loss carried out of jet cone). 

The anisotropy of medium-induced partonic energy loss goes up with increasing 
collision impact parameter $b$, because the azimuthal asymmetry of the 
interaction volume gets stronger. However, the absolute value of the energy 
loss goes down with increasing $b$ due to the reduced mean path length and the 
initial energy density. The non-uniform dependence of the loss on the parton 
azimuthal angle $\varphi$ (with respect to the reaction plane) is then mapped 
onto the final parton spectra in semi-central collisions which are approximated 
well by the elliptic form~\cite{Lokhtin:2001kb,Lokhtin:2002jd}. It results in 
the elliptic anisotropy of observed high-$p_T$ hadrons and hard jets. 
Figure~\ref{fig:v2_jet} shows calculated impact parameter dependence of $v_2$ 
coefficient for jets with $E_T^{\rm jet}>100$ GeV and for inclusive charged 
hadrons with $p_T>20$ GeV$/c$ in PbPb events triggered on jets. The absolute 
values of $v_2$ for high-$p_T$ hadrons is larger that one's for jets by a 
factor of $\sim 2-3$. However, the shape of $b$-dependence of $v^{\rm h}_2$ 
and $v^{\rm jet}_2$ is similar: it increases almost linearly with the growth 
of the impact parameter $b$ and becomes a maximum at $b \sim 1.6 R_A$ (where 
$R_A$ is the nucleus radius). After that, the $v_2$ coefficients drop rapidly 
with increasing $b$: this is the domain of impact parameter values, where the 
effect of decreasing energy loss due to a reducing effective transverse size of 
the dense zone and initial energy density of the medium is crucial and not 
compensated anymore by the stronger non-symmetry of the volume.

\subsection{$P_T$-imbalance in dimuon tagged jet events}

An important probe of medium-induced partonic energy loss in ultrarelativistic 
heavy ion collisions is production of a single jet opposite to a gauge boson 
such as a prompt $\gamma$~\cite{Wang:1996yh} or a $\gamma^\star$/$Z^0$ 
decaying into dileptons~\cite{Kartvelishvili:1995fr,Srivastava:2002kg}. 
The advantage of such processes is that the mean (i.e. averaged over all 
events) initial transverse momentum of the hard jet equal to the mean 
initial/final transverse momentum of boson, and the energy lost by the parton 
in the QCD medium can be directly estimated from the observed $p_T$-imbalance 
between the jet (or leading particle in a jet) and the lepton pair. 

In the $\gamma +$jet case the main problem arises from the jet pair production 
background when a leading $\pi^0$ in the jet is misidentified as a photon. 
The ``photon isolation'' criteria usually used in pp collisions do not work 
with the same efficiency in high multiplicity heavy ion interactions. On the 
other hand, the production of jet tagged by dileptons is not affected 
significantly by backgrounds. The main background source -- correlated 
semileptonic heavy quark decays -- can be rejected using tracker information on 
the dilepton vertex position~\cite{Virdee:1019832}. The moderate statistics, 
$\sim 500-1000$ $Z^0/\gamma^\star(\rightarrow\mu^+\mu^-)+$jet events per 1 
month of LHC run with lead beams, are expected for the CMS geometrical 
acceptance and reasonable kinematic cuts~\cite{Virdee:1019832,Lokhtin:2004zb}. 

\begin{figure}
\begin{minipage}{17pc}
\includegraphics[width=17pc]{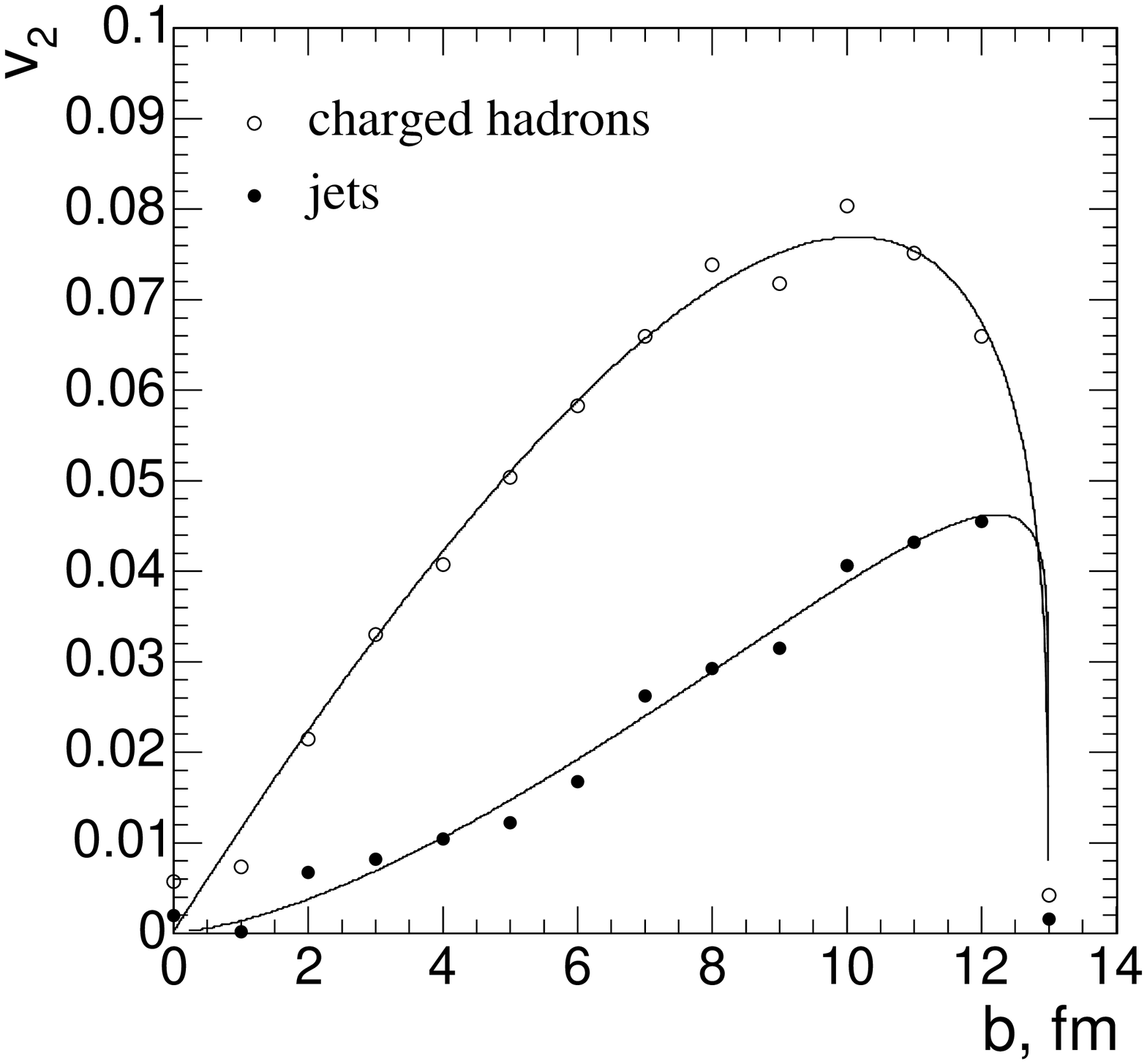}
\caption{The impact parameter dependence of elliptic flow coefficients 
$v^{\rm jet}_2$ for jets with $E_T^{\rm jet}>100$ GeV (black circles) and 
$v^{\rm h}_2$ for inclusive charged hadrons with $p_T>20$ GeV$/c$ (open 
circles) in PbPb events triggered on jets.}
\label{fig:v2_jet}
\end{minipage}
\hspace{\fill}%
\begin{minipage}{17pc}
\includegraphics[width=17pc]{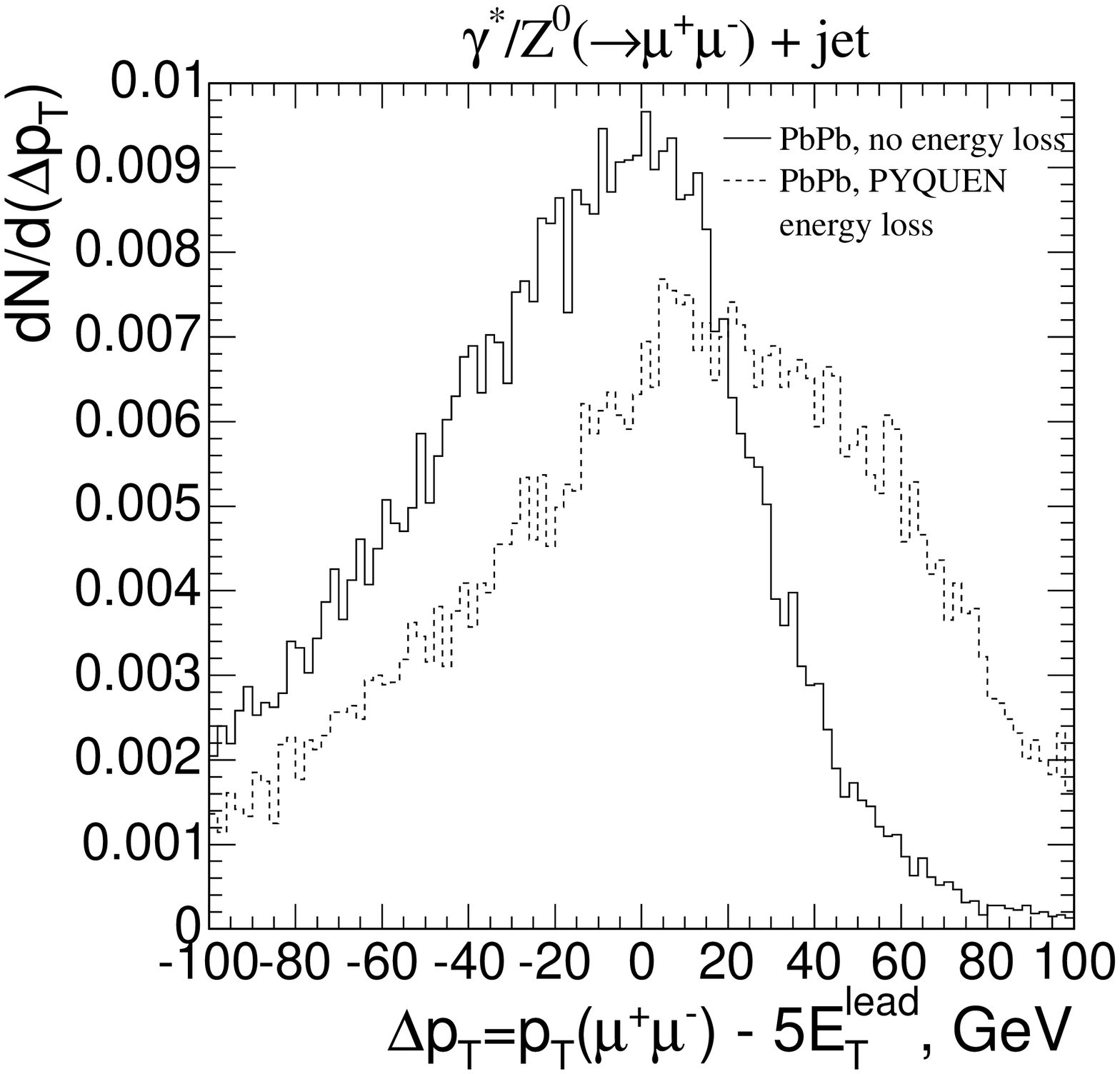}
\caption{The distribution of the difference between the transverse
momentum of a $Z^0/\gamma^\star \rightarrow\mu ^+\mu ^-$ pair,
$p_{\rm T}^{\mu ^+\mu ^-}$, and five times the transverse energy of the 
leading particle in a jet, $5 \, E_{\rm T}^{\rm lead}$, in minimum bias PbPb 
collisions with (dashed histogram) and without (solid histogram) 
partonic energy loss.}
\label{fig:mujet_elos}
\end{minipage}
\end{figure}

Figure~\ref{fig:mujet_elos} shows the difference between the transverse 
momentum of a $\mu ^+\mu ^-$ pair, $p_{\rm T}^{\mu ^+\mu ^-}$, and five times 
the transverse energy of the leading particle in a jet (since the average 
fraction of the parent parton energy carried by a leading hadron at these 
energies is $z\approx 0.2$) for minimum bias PbPb 
collisions~\cite{Lokhtin:2004zb}. The process was simulated with 
CompHEP/PYTHIA generator package without and with partonic energy loss as 
obtained in the PYQUEN.  The cuts $p_{\rm T}^{\mu}>5$ GeV$/c$, 
$p_{\rm T}^{\mu^+\mu^-}>50$ GeV$/c$ and $E_{\rm T}^{\rm jet}>50$ GeV, were  
applied. Despite the fact that the initial distribution is smeared and 
asymmetric due to initial-state gluon radiation, hadronization effects, etc., 
one can clearly see the additional smearing and the displaced mean and maximum 
values of the $p_{\rm T}$-imbalance due to partonic energy loss. The 
$p_{\rm T}$-imbalance between the $\mu ^+\mu ^-$ pair and a leading particle in 
a jet is directly related to the absolute value of partonic energy loss, and 
(unlike the $p_{\rm T}$-imbalance between the $\mu ^+\mu ^-$ pair and jet
itself) almost insensitive to the form of the angular spectrum of the emitted 
gluons or to the experimental jet energy resolution~\cite{Lokhtin:2004zb}.

\newpage

\section{Conclusions} 

The method to simulate jet quenching in heavy ion collisions has been 
developed. The model is the fast Monte-Carlo tool implemented to modify 
a standard PYTHIA jet event. The full heavy ion event is obtained as a 
superposition of a soft hydro-type state and hard multi-jets. The model is 
capable of reproducing main features of the jet quenching pattern at RHIC:  
the $p_T$ dependence of the nuclear modification factor and the suppression of 
azimuthal back-to-back correlations. The model has been applied to analyze new
features of jet quenching pattern at LHC energy: jet nuclear modification 
factor, jet fragmentation function, jet azimuthal anisotropy and dilepton-jet 
correlations. The further development of the model focusing on a more detailed 
description of low-$p_T$ particle production is in the progress. 

\section*{Acknowledgments}
 
Discussions with D.~d'Enterria, I.M.~Dremin, O.L.~Kodolova, C.~Loizides, 
A.~Morsch, L.I.~Malinina, C.~Roland, L.I.~Sarycheva, I.N.~Vardanyan, G.~Veres, 
I.~Vitev, B.~Wyslouch and B.G.~Zakharov are gratefully acknowledged. I.L. 
thanks the organizers of the Workshop for the warm welcome and hospitality.

\end{document}